\def\year{2018}\relax
\documentclass[letterpaper]{article} 
\usepackage{aaai18}  
\usepackage{times}  
\usepackage{helvet}  
\usepackage{courier}  
\usepackage{url}  
\usepackage{graphicx}  
\usepackage{natbib}

\usepackage{epsfig}
\usepackage{pgfplotstable}
\usepackage{pgfplots}
\usepackage{amsmath}
\usepackage{float}
\usepackage{hyperref}
\usepackage{microtype}
\usepackage{siunitx}
\usepackage{booktabs}
\usepackage{xcolor,colortbl}
\usepackage{multirow}
\usepackage{verbatim}
\usepackage[T1]{fontenc}
\usepackage[title]{appendix}
\usepackage{seqsplit}
\usepackage{adjustbox} 
\usepackage{todonotes}

\usepackage[utf8]{inputenc}

\usepackage{hyperref}
\hypersetup{breaklinks=true, colorlinks=true}
\usepackage{breakurl}

\begin{document}
\setcounter{secnumdepth}{2}

\title{Growing and Retaining AI Talent for the United States Government}

\author{Edward Raff\\
Booz Allen Hamilton\\
}

\maketitle

\begin{abstract}
Artificial Intelligence and Machine Learning have become transformative to a number of industries, and as such many industries need for AI talent is increasing the demand for individuals with these skills. This continues to exacerbate the difficulty of acquiring and retaining talent for the United States Federal Government, both for its direct employees as well as the companies that support it. We take the position that by focusing on growing and retaining current talent through a number of cultural changes, the government can work to remediate this problem today. 
\end{abstract}

\section{Introduction}

As Artificial Intelligence (AI) and Machine Learning (AI) become increasingly important tools for streamlining existing processes and enabling new capabilities, the United States Federal Government's demand for these skills and capabilities only increases. The standard operating procedures of most agencies within the government make attraction and retention of individuals with these skill sets difficult. 

Compensation is one significant roadblock to attracting initial talent. The average "Data Scientist" job nation wide pays \$120,931 a year\footnote{\url{https://www.glassdoor.com/Salaries/data-scientist-salary-SRCH_KO0,14.htm}}, which would be the same salary as a Step 5 GS-15 employee working for the base General Schedule \footnote{\url{https://www.opm.gov/policy-data-oversight/pay-leave/salaries-wages/salary-tables/pdf/2018/GS.pdf}}. This would require hiring staff into what is normally a senior level position with no room for future promotions, and almost no room for future increases in salary. This issue becomes more challenging when employees in this space are attracted to startups for their reduced bureaucracy, increased autonomy, and a further 10\% salary premium compared to larger organizations \cite{Kim2018}. 
While exemptions to the GS pay scale are possible, the process needed to obtain such exemptions means that they are intrinsically limited. Thus compensation will remain a major competitive disadvantage for the U.S.G. when competing for talent. 

Not only must the government work to retain its own talent, it also needs to work with the contracting agencies that supplement it's staff. While contractors can receive larger salaries than their U.S.G. employee counterparts, retention is still problematic with high demand. Potential deficiencies in the U.S.G. space that make it difficult to retain skilled staff then intrinsically impact the contracting agencies ability to retain the same skill sets when working for the government. 


We take the position that it is possible to grow and retain more AI talent within the federal government  space in the current competitive environment, provided some changes are made. We focus on changes that we believe are more realistic and obtainable for many leaders within the government, subject to the unique restrictions each may face. Namely, we recommend an approach of identifying and supporting AI "champions" with increased autonomy and support, pushing a culture of better intermingling of direct and contract employees, ensuring staff promotions do not leave "AI Vacuums", and increasing active collaborations with academia. 

\section{Growing Talent with AI Champions} \label{sec:grow_talent}

It is unlikely that the pay-scale and other systematic issues that make attracting AI talent difficult will be resolved in the near future. \citet{Kundra2010} laid out a 25-point strategy to reform the U.S.G.'s approach to Information and Technology Management, including dedicated career paths for IT management, working with congress to improve budgeting flexibility, and avoiding immutable "Grand Plan" design approaches. Eight years later many of these suggestions have not been fully realized, and a number require support from congress to make happen. 

Thus, we propose that the government focus more on growing AI talent internally. This puts more of the control within the hands of Technical Directors, and Agencies to act within their own organization to fill their needs. 

In particular, if managers can identify a AI \textit{champion} who can help shape and lead execution on mission goals, as well as grow the talent within the organization. It is crucial that this AI champion be given respect, breadth in autonomy, the freedom to investigate problems before providing a conclusion, the freedom to design solutions as they see fit to fill a need, and the ability to say "no". These all relate to problems \citet{Leetaru2016} identified as common problems preventing the effective use of data science and AI in the government.

\subsection{Supporting AI Champions}

\citet{Li2014} identified general categories of \textit{support}, \textit{ownership}, and \textit{purpose} as being necessary to obtaining and retaining talent. We argue that support is one area that many departments within the U.S.G. currently could improve. This includes support in the form of resources  (having the right hardware  and software), and education. 

\subsubsection{Compute Support}

In terms of hardware, many organizations simply do not have access to the compute resources necessary. Especially when all staff are forced to use thin-client machines which are cost effective and efficient for many purposes, but not the often compute intensive needs of AI and ML. Organizations need to be prepared to buy significant compute power for their staff if they wish to make effective use of AI, and should instead postpone their plans until funding for compute resources can be obtained. 

Once such funding exists, they should rely on their AI champion to define the compute needs and integrate them with the procurement process. In our experience, many organizations treat every dollar spent on computer equipment as equal. In reality, differing vendors may have significant changes in price for otherwise equivalent hardware. In addition, different algorithms may perform best on different types of hardware. Even simple choices such as a trade-off between more CPU cores or more RAM can be problem and algorithm specific as to what is best for the team and mission. 

It is also important to recognize that a balance in on-premise hardware and cloud compute resources is likely to exist. There are unique restrictions that can be imposed by a government agency's missions that necessitate the consideration of one of these sources in particular. When there is freedom to choose, we recommend that cloud compute be used as as the primary compute source until an AI champion can be found to help determine  the path forward. Cloud compute's flexibility in  provisioning and disposing of resources makes it a perfect fit when a compute strategy has not been determined, but progress still needs to be made without constraining the team to sub-optimal equipment for several years. Leveraging the different kinds of compute instances available can even help to make a hardware determination. 

\subsubsection{Educational Support}

Education is also critical for growing and retaining AI talent, and the education responsibilities can not rest solely on the shoulders of the AI champion. Classes at local universities, team working-sessions / hackathons, and conference attendance are all crucial components of this effort. Support for the latter two within the government has been depressed in recent years. 

Simple acts such as providing a group working space and food have minimal cost compared to employee salaries and overheads. Yet using the GSA's SmartPay system \cite{GsaSmartPay2015} to provide team lunches on an occasional basis appears to be a non-existent practice. This simple act can provide considerable benefit to employee moral and retention, while also allowing a dedicated form to disseminate lessons learned / knowledge within a working group. Even if SmartPay can't be approved, the management can set a culture by example of bringing in food to share during such hackathon sessions. 

Conference attendance in particular has been restricted across the government as a whole since a 2012 memo mandated reduced conference spending and increased oversight \cite{Zients2012}. A more recent memo has recently rolled back a number of these requirements in light of it hindering education and training in support of these agencies' important functions \cite{Donovan2016}, but we find many organizations remain just as risk-adverse to approve conference travel and spending.

Agencies need to work to remove this risk aversion to conference travel in order to train their staff. In particular, we note that considerable benefit could be achieved by reducing the turn-around time from request to approach such that it could be done within three months. This would allow staff to better select conferences based on the announced accepted papers, workshops, and tutorials. The workshops and tutorials in particular could be of considerable value due to their more focused nature, but they may not be known in the advanced time frame often demanded by the current conference approval system for many within the U.S.G.

\section{Make Contractors Part of the Team}

Contractors to the United States Federal Government, like contract and temporary workers in other sectors of the economy, can often feel like they are second class citizens within the organizations they support. This is normally caused by some disparate treatment perceived as unfair or unjust beyond recognizing the practical differences in employment. Problems could include preferential treatment of Government staff in working conditions, or discounting solutions proposed by contractors --- even if they are experts in their area. While this second-class citizen problem is  not true of all groups within the U.S.G., these issues are not new and treatment at client site directly impacts a contractor's desire to stay with both their client organization and their contracting company \cite{Boswell2012}. Because contractors are currently a crucial component of the U.S.G.'s workforce and ability to execute its mission, this issue should be of direct relevance and considerable importance to managers in the government. 

Integrating contractors as "part of the team" is about more than solving a second level retention issue for AI talent. If managers allow a greater intermixing of staff such that both federal employees and contractors could both be on teams lead by other employees or contractors, then a contractor could be leveraged as an AI champion as discussed in \autoref{sec:grow_talent}. 

Leveraging contractors for the source of these champions can allow the government to circumvent the pay-scale issue in attracting and retaining talent, as the contractors are not constrained by the GS pay scale system. For organizations which do not currently have any significant AI talent on staff, the contracting route allows them to leverage the flexibility of contract staffing to find the right champion that "fits" with the organization's culture and unique needs, avoiding the greater risk of hiring a new federal employee that might not work out. This can be particularly important if attempting to hire talent from Silicon Valley that may not adjust well to the unique constraints imposed by work in the government space \cite{Leetaru2016}. Further, empowering contractors with the equality and respect needed for them to make the proactive decisions and changes necessary to function as an AI champion has been found to improve the job performance and satisfaction of contractors working in the IT industry \cite{Huang2016}.  

We also note that contractors, in particular ones from larger organizations, can bring with them an additional social network that can be of utility. Through their parent company, the contractor may be able to reach out to or discover others within the federal government with similar needs, have previously encountered similar problems, or have compute resources they are willing to share. Leveraging both the client organization's network and the contractor's network can lead to faster results. 

We note that this kind of potential knowledge transfer can include information about tangential issues, such as how to import or export analytic code, that is important for getting work done but may not be directly about a particular AI challenge. The contractor's network may also be effective for sharing information across agencies that encounter similar problems, but may not have regular communication or even be aware that both groups are tackling the same issue.

\section{Top-Down Leadership from the Bottom-Up}

The United States government currently lacks top-down leadership in the AI space. A symptom of this is the lack of a
nation AI or ML strategy, despite the U.S. being dominant in the field as a whole. At the same time South Korea, Japan, China, the United Kingdom, and Canada have already released national strategies with other countries actively developing strategies \cite{Carter2018}. The United Arab Eremites has not only released a strategy, but created a Sate Minister for Artificial Intelligence and is pursuing AI as  integral to the government's mission of improving the quality of life for its citizens \cite{Halaweh2018}. 

This issue is important to retention as it means many organizations lack a transformational leader \cite{Bass1990} to help attract and retain talent and also improve productivity through the positive effects of a strong and consistent message \cite{Wright2009,Barling1996,Council2004} and supports the creativity needed to perform effective data science \cite{Cheung2011,Li2014}. 

While it will always be possible to hire outside talent to fill this AI leadership need, we believe much of this leadership could come from promotions of current staff. Given the large potential benefits in applying AI successfully to government missions means the success that occur at a local level from fostering AI talent could be high-profile boosts to a career. This creates the potential for developing this leadership from the bottom-up, but also requires consideration.

Individuals promoted need to work with their existing management and colleagues to coach and train their replacements. This ensures that the culture and talent developed are not transient with the manager's presence, but lasting components of the institution. If a promoted individual's replacement does not share or is not capable of continuing their mission of fostering AI talent, the staff that developed such talents will be at increased risk of leaving and creating an "AI Vacuum". Staff might leave to follow their former manager, or be lured to higher paying positions in industry when job satisfaction decreases. 

The essence of this consideration is to recognize that those who grow into AI leaders in the government are not fungible. Moving or promoting them without consideration may lead to talent loss or movement that hampers an organization and reduces productivity. This does not mean that such staff should not be promoted (indeed, we are arguing that their promotions will drive increased and wider AI talent growth!), but that supporting them includes encouraging them to identify and train their eventual successors. 

\section{Collaborate with Academia}

The need to reduce communication barriers and "stove-pipe" or "silos" within the Government has been long recognized, and duplicated efforts account for hundreds of millions in excess expenditure \cite{Dodaro2018}. Beyond wastefulness, it  can also hinder the government's goals. For example then senior CIA officer \citet{Kindsvater2003} discussed how the stove-piping in the Intelligence Community (IC) was inhibiting mission progress, and would become a larger problem as the IC's missions required more advanced and complex technology. 

This is a long standing issue that is unlikely to be resolved by managers today, and does impact the ability to retain and recruit AI talent by interfering with collaboration, creativity, and general employee happiness \cite{Leetaru2016}. For this reason we would encourage agencies to reach outside to university research groups as alternative collaboration partners to form symbiotic relationships.

For the university group, the problems faced by the U.S.G. provide real world grounded needs that make for more compelling research and publications. Individual agencies, subject to individual circumstance, may be able to share unique data that enables the research and would not be possible without the government's assistance. Some organizations within the government may in addition be able to share compute resources that would be a significant augmentation or dwarf those available to smaller research labs. 

For the government, the research group provides an augmentation of effective staff. Financially supporting a lab through a year of collaboration can be especially cost effective for the amount of work produced and time that graduate students may spend on the problem. The professor leading the research group can act as an important  source of expert AI knowledge that would be challenging to retain as an employee, but can still help current employees grow in their abilities. When results are published, the paper provides a mechanism of bridging the silo-gap by connecting with others in the government attending the same venue that the paper is published in. Finally, the connection to both graduate and undergraduate students in a lab can create a recruiting pipeline of talent that could be hired in a few years time.

We emphasize though the importance of making the relationship an  \textit{active} collaboration, with employees working with students, in order to maximize the benefits. Simply sponsoring research may help solve some problem, but fail to realize the numerous possible ancillary benefits.

\section{Conclusion}

We have discussed a number of ways in which managers within the United States Government could make changes to help grow and retain AI talent. It is unlikely that any individual will be able to apply all recommendations at once, but we believe we have outlined a set that can be practically implemented. Judgments based on individual organizations and position will determine which recommendations can and should be achieved. 

\bibliography{Mendeley}

\bibliographystyle{aaai}

\end{document}